\documentclass[twocolumn,floatfix]{revtex4}

\usepackage{graphics}

\begin{document}

\title{Emission of photon pairs at discontinuities of nonlinearity in spontaneous parametric down-conversion}

\author{Jan Pe\v{r}ina Jr., Anton\'{\i}n Luk\v{s}, Ond\v{r}ej Haderka}
\affiliation{Joint Laboratory of Optics of Palack\'{y} University
and Institute of Physics of Academy of Sciences of the Czech
Republic, 17. listopadu 50A, 772 07 Olomouc, Czech Republic}

\begin{abstract}
In order to fulfil the continuity requirements for electric- and
magnetic-field amplitudes at discontinuities of $ \chi^{(2)} $
nonlinearity additional photon pairs have to be emitted in the
area of discontinuity. Generalized two-photon spectral amplitudes
can be used to describe properties of photon pairs generated in
this process that we call surface spontaneous parametric
down-conversion. The spectral structure of such photon pairs is
similar to that derived for photon pairs generated in the volume.
Surface and volume contributions to spontaneous down-conversion
can be comparable as an example of nonlinear layered structures
shows.
\end{abstract}

\pacs{42.65.-k,42.50.-p,42.50.Dv}
\keywords{surface parametric down-conversion, surface
nonlinearity, entangled photon pair, photonic-band-gap structure}

\maketitle

\section{Introduction}

The generation of second-harmonic field at a boundary between two
homogeneous media with different values of $ \chi^{(2)} $
nonlinearity has been addressed for the first time more than 30
years ago \cite{Bloembergen1962,Bloembergen1969}. This weak effect
has been discovered when second-harmonic generation with
considerable phase mismatch has been investigated. The surface
second-harmonic field occurs here naturally and assures the
fulfilment of continuity requirements for the tangential
components of electric- and magnetic-field vector amplitudes that
stem from Maxwell's equations. In more detail, a fundamental field
creates a nonlinear polarization at second-harmonic frequency at
the nonlinear side of the boundary. This polarization generates
two surface second-harmonic fields, one in forward direction, one
in backward direction. As a consequence two different
second-harmonic fields propagate inside a nonlinear crystal. They
differ in their wave vectors. The first (and usual) field
originates from the volume nonlinear polarization and its local
wave vector is twice the wave vector of the fundamental field. On
the other hand the wave vector given by index of refraction at
second-harmonic frequency characterizes the second-harmonic field
arising at the boundary and propagating freely through the
crystal. Experimental evidence of these effects can be found,
e.g., in \cite{Bloembergen1969}. Surface second-harmonic
generation pumped by ultrashort pulses has been analyzed in
\cite{Mlejnek1999}. Deep understanding of this effect can be
reached when studying this process in a nonlinear medium with
negative index of refraction \cite{Roppo2007}. In Ref.
\cite{Roppo2007}, completely numeric approach based on the
solution of nonlinear Maxwell's equations has been adopted
contrary to the original analytical and approximate approach in
\cite{Bloembergen1962} demonstrating the richness of physical
effects included implicitly in Maxwell's equations. We note that
also inhibition of absorption in highly phase-mismatched volume
second-harmonic generation has been observed \cite{Centini2008}.

The above described effects are valid for nonlinear parametric
(three-mode) interactions in which a large number of material
states far from resonance participate. On the other hand resonant
second-harmonic generation meadiated by resonant surface states
has been widely studied for many materials (see, e.g., in
\cite{Mendoza1996,Mendoza1998}) and has become a useful tool for
surface diagnostics at present. We note that also entangled photon
pairs generated in parametric down-conversion can be converted
resonantly into plasmons at material surfaces even with the
preservation of polarization entanglement \cite{Altewischer2002}.

The question arises whether nonresonant surface effects can occur
also in the quantum process of spontaneous parametric
down-conversion (SPDC) \cite{Mandel1995}. In the case of
second-harmonic generation, the presence of macroscopic classical
nonlinear polarization at frequency $ 2\omega $ is crucial. On the
other hand, there is no macroscopic classical nonlinear
polarization at the signal- or idler-field frequencies. However,
quantum nonlinear polarization occurs in SPDC at these frequencies
and is responsible for photon-pair generation at a boundary.

SPDC is described by an appropriate momentum operator that is
constructed in the framework of energy-flux quantization
\cite{Huttner1990,Luks2002,Vogel2001}. We note that the inclusion
of all fields occurring during the propagation, i.e. forward- as
well as backward-propagating fields, is necessary to keep
consistency of the approach.

The article is divided as follows. A model of surface SPDC in case
of a homogeneous nonlinear crystal is developed in Sec.~II.
Determination of physical quantities characterizing photon pairs
is described in Sec.~III using generalized two-photon amplitudes.
Sec.~IV gives a generalization to nonlinear layered structures.
Conclusions are drawn in Sec.~V.

\section{Momentum operator and  fields' continuity at the
boundaries of a nonlinear crystal}

In this section, we first pay attention to the volume nonlinear
interaction, then study the problem at the input and later at the
output boundaries and finally add the obtained expressions
describing the photon-pair generation. We note that a simplified
model has been presented in \cite{PerinaJr2009}.

\subsection{Volume interaction}

The following interaction momentum operator $ \hat{G}_{\rm int} $
is appropriate for the process of SPDC
\cite{Mandel1995,Huttner1990}:
\begin{eqnarray}   
 \hat{G}_{\rm int}(z) &=& 4 \epsilon_0  {\cal A} \int dt
  \sum_{\alpha,\beta,\gamma=F,B} d_{\gamma,\alpha\beta} \nonumber \\
 & & \hspace{-1cm} \times \left[ E^{(+)}_{p_\gamma}(z,t)
  \hat{E}^{(-)}_{s_\alpha}(z,t) \hat{E}^{(-)}_{i_\beta}(z,t)
  + {\rm h.c.} \right] ,
\label{1}
\end{eqnarray}
where $ E^{(+)}_{p_\gamma} $ are positive-frequency parts of the
(linearly polarized) pump-field electric-field amplitudes whereas
$ \hat{E}^{(-)}_{s_\alpha} $ ($ \hat{E}^{(-)}_{i_\beta} $) stand
for negative-frequency parts of the signal- (idler-) field
electric-field amplitude operators. Subscript $ F $ ($ B $) refers
to a field propagating forward (backward), i.e. along the $ +z $
($ -z $) axis. Symbol $ \epsilon_0 $ means permittivity of vacuum,
$ d_{\gamma,\alpha\beta} $ are effective nonlinear coefficients, $
{\cal A} $ is transverse area of the fields, and $ {\rm h.c.} $
replaces the hermitian-conjugated terms. We have assumed a scalar
model for the interacting fields for simplicity. However, a
generalization to the vectorial model is straightforward because
of the applied first-order perturbation approximation. Using a
spectral decomposition of the interacting fields,
\begin{eqnarray}  
 E_{m_\alpha}^{(+)}(z,t) &=& \frac{1}{\sqrt{2\pi}} \int d\omega_m \,
  E_{m_\alpha}^{(+)}(z,\omega_m) \exp(-i\omega_m t) , \nonumber \\
   & & \hspace{2cm} m = p,s,i, \;\; \alpha= F,B,
\label{2}
\end{eqnarray}
the momentum operator $ \hat{G}_{\rm int} $ in Eq.~(\ref{1}) can
be recast into the form
\begin{eqnarray}   
 \hat{G}_{\rm int}(z) &=& \frac{4 \epsilon_0 {\cal A}}{ \sqrt{2\pi}}
  \int d\omega_p \int d\omega_s \int d\omega_i \, \delta(\omega_p-\omega_s-\omega_i)
  \nonumber \\
 & &
  \hspace{-15mm}  \times \sum_{\alpha,\beta,\gamma=F,B}
  d_{\gamma,\alpha\beta} \nonumber \\
 & & \hspace{-15mm} \mbox{} \times
  \left[ E^{(+)}_{p_\gamma}(z,\omega_p) \hat{E}^{(-)}_{s_\alpha}(z,\omega_s)
  \hat{E}^{(-)}_{i_\beta}(z,\omega_i)
  + {\rm h.c.} \right] ,
\label{3}
\end{eqnarray}
where the $ \delta $-function expresses conservation of energy for
monochromatic waves. The signal and idler spectral electric-field
amplitude operators $ \hat{E}^{(-)}_{m_\alpha} $ can be expressed
in terms of creation operators $ \hat{a}_{m_\alpha}^\dagger $
introduced such that $ \hat{a}^\dagger_{m_\alpha}
\hat{a}_{m_\alpha} $ gives the photon-number density in mode $
m_\alpha $ at a given frequency:
\begin{eqnarray} 
 \hat{E}_{m_\alpha}^{(-)}(z,\omega_m) &=& - i \sqrt{ \frac{
  \hbar\omega_m }{ 2\epsilon_0 c{\cal A} n_m(\omega_m) } }
  \hat{a}^\dagger_{m_\alpha}(z,\omega_m), \nonumber \\
  & & \hspace{10mm} m = s,i, \;\; \alpha=F,B.
\label{4}
\end{eqnarray}
Symbol $ n_m $ stands for an index of refraction of field $ m $.

Spatial evolution of optical fields is determined by the solution
of Heisenberg equations \cite{Perina1991,PerinaJr2000} for field
operators denoted as $ \hat{X} $:
\begin{eqnarray}   
 \frac{d\hat{X}(z)}{dz} &=& - \frac{i}{\hbar} \left[
  \hat{G}(z),\hat{X}(z) \right] ; \\
 \hat{G}(z) &=& \hat{G}_0(z) + \hat{G}_{\rm int}(z) , \nonumber \\
 \hat{G}_0(z) &=&  \sum_{m=s,i} \sum_{\alpha=F,B} \hbar \int d\omega_m k_{m_\alpha}(\omega_m)
  \nonumber \\
 & & \times \hat{a}_{m_\alpha}^\dagger(z,\omega_m) \hat{a}_{m_\alpha}(z,\omega_m) ;
\label{6}
\end{eqnarray}
the interaction momentum operator $ \hat{G}_{\rm int} $ is given
in Eq.~(\ref{3}). The momentum operator $ \hat{G}_0 $ introduced
in Eq.~(\ref{6}) describes free-field evolution. Symbol $
k_{m_\alpha} $ stands for a wave vector of mode $ m_\alpha $ at
frequency $ \omega_m $; $ k_{m_F} = k_m $, $ k_{m_B} = - k_m $, $
k_m > 0 $. Symbol $ \hbar $ denotes the reduced Planck constant.

In order to determine the electric-field amplitude operators at
the output of the nonlinear crystal, the Heisenberg equations for
operators $ \hat{a}_{m_\alpha}(z,\omega_m) $ ($ m=s,i $, $
\alpha=F,B $) have to be solved:
\begin{eqnarray}    
 \frac{ d\hat{a}_{s_\alpha}(z,\omega_s) }{dz} &=& i k_{s_\alpha}(\omega_s)
  \hat{a}_{s_\alpha}(z,\omega_s)  \nonumber \\
 & & \hspace{-2cm} +  \sum_{\beta,\gamma=F,B}
  \int d\omega_i \, g_{\gamma,\alpha\beta} (\omega_s,\omega_i) E^{(+)}_{p_\gamma}(0,\omega_s+\omega_i)
  \nonumber \\
 & & \hspace{-2cm} \times \exp[ik_{p_\gamma}(\omega_s+\omega_i)z]
  \hat{a}_{i_\beta}^\dagger(z,\omega_i), \;\; \alpha=F,B.
\label{7}
\end{eqnarray}
Equations (\ref{7}) have been derived assuming equal-space
commutation relations \cite{PerinaJr2000}. We note that fields
propagating along the $ -z $ axis have negative wave-vectors in
the definition of the free-field momentum operator $ \hat{G}_0 $
in Eq.~(\ref{6}). A more detailed and rigorous formulation of the
dynamics of counter-propagating fields justifying this approach
can be found in \cite{PerinaJr2000}. The coupling constants $
g_{\gamma,\alpha\beta} $ occurring in Eq.~(\ref{7}) are given
along the expression:
\begin{eqnarray}  
 g_{\gamma,\alpha\beta}(\omega_s,\omega_i) &=& \frac{2i d_{\gamma,\alpha\beta} }{c}
  \sqrt{ \frac{\omega_s \omega_i}{2\pi n_s(\omega_s) n_i(\omega_i)}
  } .
\label{8}
\end{eqnarray}
Equations for operators $ \hat{a}_{i_F} $ and $ \hat{a}_{i_B} $
can be derived from Eq.~(\ref{7}) by the formal substitution $ s
\leftrightarrow i $.

Solution of Eq.~(\ref{7}) valid up to the first power of $ g $ can
be obtained in the form:
\begin{eqnarray}  
  \hat{a}_{s_\alpha}(z,\omega_s) &=& \exp[ ik_{s_\alpha}(\omega_s) z ]
  \left[ \hat{a}_{s_\alpha}(0,\omega_s) \right.  \nonumber \\
 & & \hspace{-3cm} \left. +  \sum_{\beta,\gamma=F,B}
  \int d\omega_i
  {\cal B}_{\gamma,\alpha\beta}(z,\omega_s,\omega_i)
  \hat{a}_{i_\beta}^\dagger(0,\omega_i) \right], \nonumber \\
 & & \hspace{1cm}  \alpha=F,B,
\label{9}
\end{eqnarray}
where
\begin{eqnarray}  
 {\cal B}_{\gamma,\alpha\beta}(z,\omega_s,\omega_i) &=& g_{\gamma,\alpha\beta}
   (\omega_s,\omega_i)
   E^{(+)}_{p_\gamma}(0,\omega_s+\omega_i) \nonumber \\
 & & \times \exp[i\Delta k_{\gamma,\alpha\beta}(\omega_s,\omega_i)
   z/2] \nonumber \\
 & &  \times  z \,
   {\rm sinc} [\Delta k_{\gamma,\alpha\beta}(\omega_s,\omega_i) z/2] ;
\label{10}  \\
 \Delta k_{\gamma,\alpha\beta}(\omega_s,\omega_i) &=& k_{p_\gamma}(\omega_s+\omega_i) -
  k_{s_\alpha}(\omega_s) - k_{i_\beta}(\omega_i) ; \nonumber \\
 & & \hspace{1cm} \alpha,\beta,\gamma=F,B;
\label{11}
\end{eqnarray}
$ {\rm sinc}(x) = \sin(x)/x $. We note that the restriction to the
first power of $ g $ in the formula in Eq.~(\ref{9}) has justified
the use of the approximate formula $
\hat{a}_{m_\alpha}(z,\omega_m) = \exp[ ik_{m_\alpha}(\omega_m) z ]
\hat{a}_{m_\alpha}(0,\omega_m) $ below the integral over the
frequency $ \omega_i $. The solution in Eq.~(\ref{9}) describes
SPDC originating in the volume of nonlinear crystal.

The solution obtained for annihilation operators $
\hat{a}_{s_\alpha}(z,\omega_s) $ as written in Eq.~(\ref{9})
provides the following expressions for the positive-frequency
parts of electric- ($ \hat{E}^{(+)}_{s_\alpha} $) and
magnetic-field ($ \hat{H}^{(+)}_{s_\alpha} $) amplitude operators:
\begin{eqnarray}   
 \hat{E}_{s_\alpha}^{(+)}(z,\omega_s) &=& i \sqrt{ \frac{
  \hbar\omega_s}{ 2\epsilon_0 c {\cal A} n_s(\omega_s) } }
  \exp[ ik_{s_\alpha}(\omega_s) z ]
  \nonumber \\
 & & \hspace{-1cm} \Biggl[ \hat{a}_{s_\alpha}(0,\omega_s) +
  \sum_{\beta,\gamma=F,B} \nonumber \\
 & & \hspace{-1cm}
  \int d\omega_i {\cal B}_{\gamma,\alpha\beta}(z,\omega_s,\omega_i)
  \hat{a}_{i_\beta}^\dagger(0,\omega_i) \Biggr], \nonumber \\
 & &
\label{12}  \\
 \hat{H}_{s_\alpha}^{(+)}(z,\omega_s) &=&
  \hat{H}_{s_\alpha}^{(+){\rm Fr}}(z,\omega_s) +
  \hat{H}_{s_\alpha}^{(+){\rm nFr}}(z,\omega_s), \nonumber \\
 & &
\label{13} \\
 \hat{H}_{s_\alpha}^{(+){\rm Fr}}(z,\omega_s) &=&
 \frac{k_{s_\alpha}(\omega_s)}{\omega_s\mu_0}
 \hat{E}_{s_\alpha}^{(+)}(z,\omega_s),
\label{14} \\
 \hat{H}_{s_\alpha}^{(+){\rm nFr}}(z,\omega_s) &=& \sqrt{ \frac{ \hbar c
  }{ 2\mu_0 \omega_s {\cal A} } n_s(\omega_s) } \nonumber \\
 & & \hspace{-2cm} \times \sum_{\beta,\gamma=F,B} \int d\omega_i\, g_{\gamma,\alpha\beta}
  (\omega_s,\omega_i)
  E^{(+)}_{p_\gamma}(\omega_s+\omega_i) \nonumber \\
  & & \hspace{-2cm} \times \exp[ik_{p_\gamma}(\omega_s+\omega_i) z] \exp[-ik_{i_\beta}(\omega_i)z]
  \hat{a}^\dagger_{i_\beta}(0,\omega_i) , \nonumber \\
 & & \hspace{25mm} \alpha=F,B .
\label{15}
\end{eqnarray}
Equations (\ref{13}---\ref{15}) for the magnetic-field amplitude
operators $ \hat{H}_{s_\alpha} $ ($ \alpha=F,B $) have been
derived assuming polarization of electric-field amplitudes $
\hat{E}_{s_\alpha} $ along the $ +x $ axis and, consequently,
polarization of magnetic-field amplitudes $ \hat{H}_{s_\alpha} $
along the $ +y $ axis. The Maxwell equations then provide the
following formula $ H_{s_\alpha}^{(+)}(z,\omega_s) = -
i/(\omega_s\mu_0)
\partial E_{s_\alpha}^{(+)}(z,\omega_s) / \partial z $, where $ \mu_0 $
stands for permeability of vacuum. The magnetic-field amplitude
operators $ \hat{H}^{(+)}_{s_\alpha}(z,\omega_s) $ have been
decomposed in Eq.~(\ref{13}) into two parts; the amplitude
operators $ \hat{H}^{(+){\rm Fr}}_{s_\alpha}(z,\omega_s) $ are
linearly proportional to the electric-field amplitude operators $
\hat{E}^{(+)}_{s_\alpha}(z,\omega_s) $ whereas the amplitude
operators $ \hat{H}^{(+){\rm nFr}}_{s_\alpha}(z,\omega_s) $ are of
purely nonlinear origin. The amplitude operators $
\hat{H}^{(+){\rm nFr}}_{s_\alpha} $ are not taken into account in
the usual derivation of Fresnel's relations that assumes linear
media. Correct inclusion of these amplitude operators into the
continuity considerations at a boundary results in additional
contributions to the nonlinear process.

\subsection{Input boundary}

\begin{figure}         
 \resizebox{0.98\hsize}{!}{\includegraphics{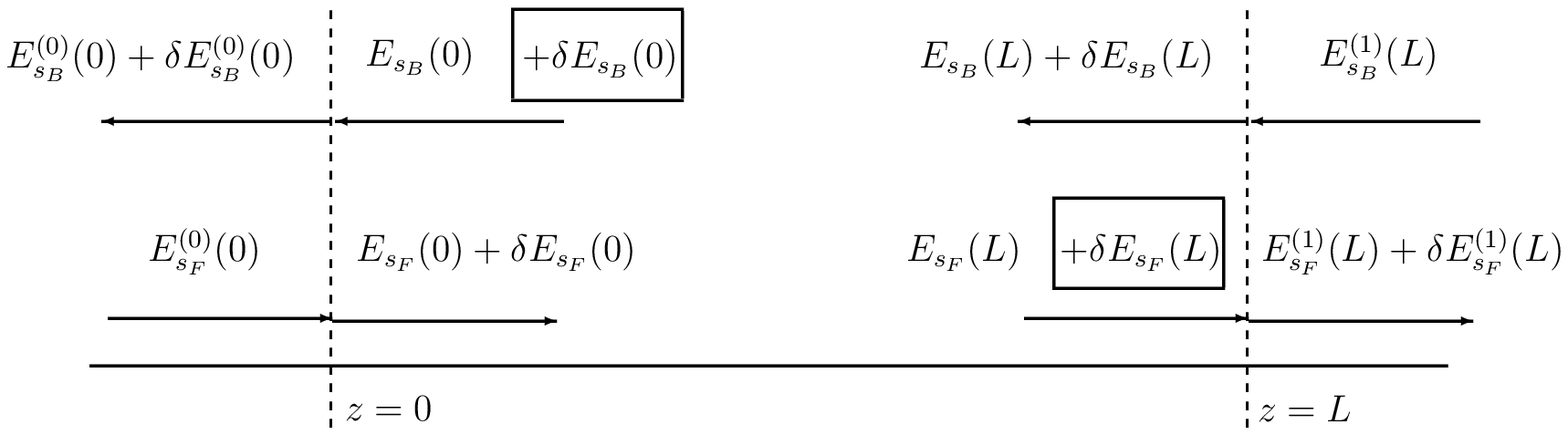}}
  \caption{Scheme showing electric-field amplitudes $ E $ and their surface corrections
  $ \delta E $ at the input ($ z=0 $) and output ($ z=L $) boundaries of
  a nonlinear crystal. Superscript $ (0) $ [$ (1) $] denotes amplitudes in front
  [beyond] the nonlinear crystal. Amplitude corrections $ \delta E_{s_B}(0) $ and $ \delta
  E_{s_F}(L) $ written in frame-boxes do not exist in the real nonlinear medium; they replace the
  effect of real amplitude corrections $ \delta E_{s_F}(0) $, $ \delta
  E_{s_B}^{(0)}(0) $, $ \delta E_{s_F}^{(1)}(L) $, and $ \delta E_{s_B}(L) $.}
\label{fig1}
\end{figure}

Let us take deeper attention to the problem of continuity of
electric- and magnetic-field amplitudes at the boundaries of the
nonlinear medium. First we pay attention to the input boundary ($
z=0 $) and consider the signal field. Four electric- and
magnetic-field amplitudes are involved in the continuity
requirement: amplitudes $ E_{s_F}^{(0)}(0) $ and $
H_{s_F}^{(0)}(0) $ of the forward-propagating field impinging on
the boundary from the outside of nonlinear medium, amplitudes $
E_{s_B}^{(0)}(0) $ and $ H_{s_B}^{(0)}(0) $ leaving the boundary
outside the nonlinear crystal, amplitudes $ E_{s_B}(0) $ and $
H_{s_B}(0) $ of the field coming to the boundary from the
nonlinear crystal, and amplitudes $ E_{s_F}(0) $ and $ H_{s_F}(0)
$ leaving the boundary and propagating inside the nonlinear
crystal [see Fig.~(\ref{fig1})]. Because the magnetic-field
amplitudes $ H_{s_F}^{}(0) $ and $ H_{s_B}^{}(0) $ defined in the
nonlinear crystal have also nonlinear contributions $ H_{s_F}^{\rm
nFr}(0) $ and $ H_{s_B}^{\rm nFr}(0) $ described in Eq.~(\ref{15})
additional (surface) amplitude corrections $ \delta E_{s_F}(0) $
and $ \delta E_{s_B}^{(0)}(0) $ together with $ \delta H_{s_F}(0)
$ and $ \delta H_{s_B}^{(0)}(0) $ are needed to fulfil the
continuity requirement. The surface amplitude corrections
naturally occur in the fields that leave the boundary which is a
consequence of spatio-temporal considerations that are suppressed
to certain extent in our one-dimensional model.

The requirement of continuity of projections of electric- and
magnetic-field amplitudes to the plane of the input boundary leads
to the following equations:
\begin{eqnarray}   
 E_{s_F}^{(0)}(0) + E_{s_B}^{(0)}(0) + \delta E_{s_B}^{(0)}(0) &=&
  E_{s_F}(0) + \delta E_{s_F}(0) \nonumber \\
 & & \hspace{-3cm} + E_{s_B}(0) ,
\label{16} \\
 H_{s_F}^{(0)}(0) + H_{s_B}^{(0)}(0) + \delta H_{s_B}^{(0)}(0) &=&
  H_{s_F}^{\rm Fr}(0) + H_{s_F}^{\rm nFr}(0) \nonumber \\
 & & \hspace{-3cm} + \delta H_{s_F}(0) +
 H_{s_B}^{\rm Fr}(0) + H_{s_B}^{\rm nFr}(0) .
\label{17}
\end{eqnarray}
Derivation of the usual Fresnel's relations (in linear media)
\cite{Wolf1980} is based on the fulfilment of the following
equations:
\begin{eqnarray}   
 E_{s_F}^{(0)}(0) + E_{s_B}^{(0)}(0)  &=&
 E_{s_F}(0) + E_{s_B}(0) ,
\label{18} \\
 H_{s_F}^{(0)}(0) + H_{s_B}^{(0)}(0) &=&
 H_{s_F}^{\rm Fr}(0) + H_{s_B}^{\rm Fr}(0).
\label{19}
\end{eqnarray}
Comparison of Eqs.~(\ref{16}) and (\ref{18}) together with
Eqs.~(\ref{17}) and (\ref{19}) results in two algebraic equations
for the surface amplitude corrections of the fields leaving the
boundary:
\begin{eqnarray}   
 \delta E_{s_B}^{(0)}(0) &=& \delta E_{s_F}(0) ,
\label{20} \\
 \delta H_{s_B}^{(0)}(0) &=& H_{s_F}^{nFr}(0) + \delta H_{s_F}(0) +
  H_{s_B}^{\rm nFr}(0) .
\label{21}
\end{eqnarray}

Alternatively and more conveniently, the amplitude corrections $
\delta E_{s_B}^{(0)} $ and $ \delta H_{s_B}^{(0)} $ of the field
outside the nonlinear crystal can be formally included into the
equations giving Fresnel's relations. This can be done if we
introduce fictitious amplitude corrections $ \delta E_{s_B} $ and
$ \delta H_{s_B} $ of the field impinging on the boundary from its
nonlinear side. Such corrections give, after transformation at the
boundary using Fresnel's relations, the required amplitude
corrections $ \delta E_{s_B}^{(0)} $ and $ \delta H_{s_B}^{(0)} $
[see Fig.~\ref{fig1}].
Then we have:
\begin{eqnarray}   
 E_{s_F}^{(0)}(0) + \left[ E_{s_B}^{(0)}(0) +
  \delta E_{s_B}^{(0)}(0) \right] &=&
  E_{s_F}(0) \nonumber \\
 & & \hspace{-3cm} + \left[ E_{s_B}(0) + \delta E_{s_B}(0) \right] ,
\label{22} \\
 H_{s_F}^{(0)}(0) + \left[ H_{s_B}^{(0)}(0) + \delta H_{s_B}^{(0)}(0) \right] &=&
  H_{s_F}^{\rm Fr}(0) \nonumber \\
 & & \hspace{-3cm} + \left[ H_{s_B}^{\rm Fr}(0) + \delta H_{s_B}(0) \right].
\label{23}
\end{eqnarray}
Equations (\ref{16}) and (\ref{17}) are then fulfilled provided
that the following two algebraic equations for surface amplitude
corrections of the fields inside the nonlinear crystal are valid:
\begin{eqnarray}   
 0 &=& \delta E_{s_F}(0) -  \delta E_{s_B}(0),
\label{24} \\
 0 &=& H_{s_F}^{\rm nFr}(0) + \delta H_{s_F}(0) +
  H_{s_B}^{\rm nFr}(0) - \delta H_{s_B}(0) . \nonumber \\
 & &
\label{25}
\end{eqnarray}

The positive-frequency parts of surface amplitude-correction
operators $ \delta \hat{E}_{m_\alpha} $ and $ \delta
\hat{H}_{m_\alpha} $ are defined similarly as the corresponding
amplitude operators $ \hat{E}_{m_\alpha} $ and $
\hat{H}_{m_\alpha} $ in Eqs.~(\ref{4}) and (\ref{14}) using
operator corrections $ \delta\hat{a}_{m_\alpha} $ to the
annihilation operators $ \hat{a}_{m_\alpha} $:
\begin{eqnarray}   
 \delta \hat{E}_{m_\alpha}^{(+)}(z,\omega_m) &=& i \sqrt{ \frac{
  \hbar\omega_m}{ 2\epsilon_0 c{\cal A} n_m(\omega_m) } }
  \delta\hat{a}_{m_\alpha}(z,\omega_m),
\label{26} \\
 \delta\hat{H}_{m_\alpha}^{(+)}(z,\omega_m) &=&
  \frac{k_{m_\alpha}(\omega_m)}{\omega_m\mu_0}
  \delta\hat{E}_{m_\alpha}^{(+)}(z,\omega_m) , \nonumber \\
 & & \hspace{1cm} m=s,i,\,\, \alpha=F,B. \label{27}
\end{eqnarray}

Substitution of Eqs.~(\ref{15}), (\ref{26}), and (\ref{27}) into
Eqs.~(\ref{24}) and (\ref{25}) gives two algebraic equations for
the annihilation-operator corrections $ \delta
\hat{a}_{s_F}(0,\omega_s) $ and $ \delta \hat{a}_{s_B}(0,\omega_s)
$:
\begin{eqnarray}    
 \delta\hat{a}_{s_F}(0,\omega_s) - \delta \hat{a}_{s_B}(0,\omega_s) &=& 0,
 \nonumber \\
 & &
\label{28}  \\
 ik_{s_F}(\omega_s) \delta\hat{a}_{s_F}(0,\omega_s) - ik_{s_B}(\omega_s)
  \delta \hat{a}_{s_B}(0,\omega_s)  & &  \nonumber \\
  \mbox{} + \sum_{\alpha,\beta,\gamma=F,B}  \int d\omega_i\,
  g_{\gamma,\alpha\beta}(\omega_s,\omega_i) & & \nonumber \\
  \mbox{} \times E^{(+)}_{p_\gamma}(\omega_s+\omega_i)
  \hat{a}^{\dagger}_{i_\beta}(0,\omega_i) &=& 0 . \nonumber \\
  & &
\label{29}
\end{eqnarray}

Solution of Eqs~(\ref{28}) and (\ref{29}) finally gives the
expressions for annihilation-operator corrections $ \delta
\hat{a}_{s_F} $ and $ \delta \hat{a}_{s_B} $ at the input
boundary:
\begin{eqnarray}  
 \delta a_{s_F}(0,\omega_s) &=& \frac{i}{2k_s(\omega_s)}
  \sum_{\alpha,\beta,\gamma=F,B} \int d\omega_i\,
  g_{\gamma,\alpha\beta}(\omega_s,\omega_i)
  \nonumber \\
 & & \times  E^{(+)}_{p_\gamma}(\omega_s+\omega_i) \hat{a}^\dagger_{i_\beta}(0,\omega_i)
\label{30} \\
 \delta a_{s_B}(0,\omega_s) &=& \delta a_{s_F}(0,\omega_s) .
\label{31}
\end{eqnarray}

\subsection{Output boundary}

Fields at the output boundary of the nonlinear crystal can be
analyzed similarly as at the input boundary. Here, the continuity
of projections of electric- and magnetic-field amplitudes at $ z =
L $ ($ L $ denotes the crystal length) to the plane of the
boundary gives two equations:
\begin{eqnarray}   
 & & E_{s_F}(L) + E_{s_B}(L) + \delta E_{s_B}(L) \nonumber \\
 & & \hspace{1cm} =
 E_{s_F}^{(1)}(L) + \delta E_{s_F}^{(1)}(L) + E_{s_B}^{(1)}(L) ,
\label{32} \\
 & & H_{s_F}^{\rm Fr}(L) + H_{s_F}^{\rm nFr}(L) + H_{s_B}^{\rm Fr}(L)
  + H_{s_B}^{\rm nFr}(L) + \delta H_{s_B}(L) \nonumber \\
 & & \hspace{1cm}
  = H_{s_F}^{(1)}(L) + \delta H_{s_F}^{(1)}(L) +
  H_{s_B}^{(1)}(L).
\label{33}
\end{eqnarray}
Amplitudes $ E_{s_F}^{(1)}(L) $ and $ H_{s_F}^{(1)}(L) $ describe
the field outside the nonlinear crystal whereas amplitudes $
E_{s_B}^{(1)}(L) $ and $ H_{s_B}^{(1)}(L) $ refer to the field
impinging on the output boundary from its linear side (see
Fig.~\ref{fig1}). The amplitude corrections $ \delta
E_{s_F}^{(1)}(L) $ and $ \delta H_{s_F}^{(1)}(L) $ can be formally
included into the equations that express Fresnel's relations
provided that fictitious amplitude corrections $ \delta E_{s_F}(L)
$ and $ \delta H_{s_F}(L) $ are introduced. Motivation for this
step is the same as in the case of input boundary: we want to have
corrections only inside the nonlinear crystal. We then have:
\begin{eqnarray}   
 \left[ E_{s_F}(L) + \delta E_{s_F}(L) \right] + E_{s_B}(L) &=&
  \nonumber \\
 & & \hspace{-4cm} \left[ E_{s_F}^{(1)}(L) + \delta E_{s_F}^{(1)}(L)
 \right] + E_{s_B}^{(1)}(L) ,
\label{34} \\
 \left[ H_{s_F}^{\rm Fr}(L) + \delta H_{s_F}(L) \right] + H_{s_B}^{\rm Fr}(L) &=&
 \nonumber \\
 & & \hspace{-4cm}
  \left[ H_{s_F}^{(1)}(L) + \delta H_{s_F}^{(1)}(L) \right] + H_{s_B}^{(1)}(L).
\label{35}
\end{eqnarray}
Comparison of Eqs.~(\ref{34}) and (\ref{35}) with Eqs.~(\ref{32})
and (\ref{33}) results in two algebraic equations for the surface
amplitude corrections inside the nonlinear medium:
\begin{eqnarray}   
 - \delta E_{s_F}(L) + \delta E_{s_B}(L) &=& 0 , \nonumber \\
 & &
\label{36} \\
 H_{s_F}^{\rm nFr}(L) - \delta H_{s_F}(L) + H_{s_B}^{\rm nFr}(L) + \delta H_{s_B}(L) &=&
 0 . \nonumber \\
 & &
\label{37}
\end{eqnarray}

Two algebraic equations for the annihilation-operator corrections
$ \delta \hat{a}_{s_F}(L,\omega_s) $ and $ \delta
\hat{a}_{s_B}(L,\omega_s) $ can be derived from Eqs. (\ref{36})
and (\ref{37}) using the expressions in Eqs. (\ref{15}),
(\ref{26}), and (\ref{27}):
\begin{eqnarray}    
 -\delta\hat{a}_{s_F}(L,\omega_s) + \delta \hat{a}_{s_B}(L,\omega_s) &=& 0,
  \nonumber \\
 & &
\label{38}  \\
 -ik_{s_F}(\omega_s) \delta\hat{a}_{s_F}(L,\omega_s) + ik_{s_B}(\omega_s)
   \delta \hat{a}_{s_B}(L,\omega_s) & & \nonumber \\
 + \sum_{\alpha,\beta,\gamma=F,B} \int d\omega_i\,
  g_{\gamma,\alpha\beta}(\omega_s,\omega_i) \nonumber \\
 \mbox{} \times E^{(+)}_{p_\gamma}(\omega_s+\omega_i) \exp[ik_{p_\gamma}(\omega_s+\omega_i)
   L]& &  \nonumber \\
 \mbox{} \times \exp[-ik_{i_\beta}(\omega_i)L]
  \hat{a}^\dagger_{i_\beta}(0,\omega_i) &=& 0 . \nonumber \\
 & &
\label{39}
\end{eqnarray}

Equations~(\ref{38}) and (\ref{39}) can be solved leaving us the
expressions for annihilation-operator corrections $ \delta
\hat{a}_{s_F}(L,\omega_s) $ and $ \delta \hat{a}_{s_B}(L,\omega_s)
$ at the output boundary:
\begin{eqnarray}  
 \delta a_{s_F}(L,\omega_s) &=& \frac{-i}{2k_s(\omega_s)}
  \sum_{\alpha,\beta,\gamma=F,B} \int d\omega_i\, g_{\gamma,\alpha\beta}(\omega_s,\omega_i)
  \nonumber \\
 & & \times E^{(+)}_{p_\gamma}(\omega_s+\omega_i) \exp[ik_{p_\gamma}(\omega_s+\omega_i) L]
  \nonumber \\
 & & \times \exp[-ik_{i_\beta}(\omega_i)L]
 \hat{a}^\dagger_{i_\beta}(0,\omega_i),
\label{40} \\
 \delta a_{s_B}(L,\omega_s) &=& \delta a_{s_F}(L,\omega_s) .
\label{41}
\end{eqnarray}

\subsection{The whole nonlinear crystal}

The overall solution for operators $
\hat{a}_{s_\alpha}(L,\omega_s) $ valid up to the first power of $
g $ has three nonlinear contributions: the first comes from the
input boundary, the second from the volume and the third from the
output boundary. The overall solution can be written as follows:
\begin{eqnarray}    
 \hat{a}_{s_\alpha}(L,\omega_s) &=&
   \hat{a}_{s_\alpha}^{\rm free}(L,\omega_s) + \sum_{\beta,\gamma=F,B}
   \nonumber \\
  & & \hspace{-3cm} \int d\omega_i \,
   {\cal F}^s_{\gamma,\alpha\beta}(L,\omega_s,\omega_i)
   \hat{a}_{i_\beta}^{{\rm free}\dagger}(L,\omega_i), \;\; \alpha=F,B.
\label{42}
\end{eqnarray}
Operators $ \hat{a}_{m_\alpha}^{{\rm free}}(L,\omega_m) $ ($ m=s,i
$, $ \alpha=F,B $) occurring in Eqs.~(\ref{42}) and (\ref{43})
below and expressed at the crystal end correspond to free-field
evolution (i.e., without photon-pair generation) inside the
crystal. They are determined by the formula $
\hat{a}_{m_\alpha}^{{\rm free}}(L,\omega_m) =
\exp[ik_{m_\alpha}(\omega_m)L] \hat{a}_{m_\alpha}^{{\rm
free}}(0,\omega_m) $. Functions $ {\cal
F}^{s}_{\gamma,\alpha\beta} $ are defined in
Eqs.~(\ref{44}---\ref{46}) bellow.

Now we pay attention to the idler fields and use symmetry between
the signal and idler fields. The idler-field electric- (magnetic-)
field amplitudes are assumed to be polarized along the $ +y $ ($
-x $) axis. The requirement of continuity of electric- and
magnetic-field amplitudes at the input and output boundaries leads
to the solution for idler-field operators $
\hat{a}_{i_\beta}(L,\omega_i) $ in the form:
\begin{eqnarray}    
 \hat{a}_{i_\beta}(L,\omega_i) &=&
   \hat{a}_{i_\beta}^{\rm free}(L,\omega_i) + \sum_{\alpha,\gamma=F,B}
   \nonumber \\
  & & \hspace{-3cm} \int d\omega_s \,
   {\cal F}^i_{\gamma,\alpha\beta}(L,\omega_s,\omega_i)
   \hat{a}_{s_\alpha}^{{\rm free}\dagger}(L,\omega_s), \;\; \beta=F,B.
\label{43}
\end{eqnarray}

The functions $ {\cal F}^s $ and $ {\cal F}^i $ introduced in
Eqs.~(\ref{42}) and (\ref{43}) can be decomposed into volume ($
{\cal F}^{\rm vol} $) and surface ($ {\cal F}^{s,\rm surf} $, $
{\cal F}^{i,\rm surf} $) contributions:
\begin{eqnarray}   
 {\cal F}^m_{\gamma,\alpha\beta} &=& {\cal F}^{\rm vol}_{\gamma,\alpha\beta}
  + {\cal F}^{m,\rm surf}_{\gamma,\alpha\beta},
\label{44}  \\
 {\cal F}^{\rm vol}_{\gamma,\alpha\beta}(L,\omega_s,\omega_i) &=& g_{\gamma,\alpha\beta}(\omega_s,\omega_i)
  E^{(+)}_{p_\gamma}(\omega_s+\omega_i) \nonumber \\
 & & \hspace{-3cm} \times  \exp[ik_{p_\gamma}(\omega_s+\omega_i)L]
   \exp[-i\Delta k_{\gamma,\alpha\beta}(\omega_s,\omega_i)
   L/2] \nonumber \\
 & & \hspace{-3cm} \times  L \,
   {\rm sinc} [\Delta k_{\gamma,\alpha\beta}(\omega_s,\omega_i) L/2] ,
\label{45}
  \\
 {\cal F}^{m,\rm surf}_{\gamma,\alpha\beta}(L,\omega_s,\omega_i) &=& \frac{i}{k_m(\omega_m) }
  g_{\gamma,\alpha\beta}(\omega_s,\omega_i) E^{(+)}_{p_\gamma}(\omega_s+\omega_i)
  \nonumber \\
 & & \hspace{-1.5cm} \times
  \left\{ \exp[ik_{s_\alpha}(\omega_s)L] \exp[ik_{i_\beta}(\omega_i)L] \right.
  \nonumber \\
 & & \hspace{-1.5cm} \left.  -
  \exp[ik_{p_\gamma}(\omega_s+\omega_i)L] \right\} , \nonumber \\
 & &  m=s,i; \hspace{5mm} \alpha,\beta,\gamma = F,B .
\label{46}
\end{eqnarray}
In deriving Eq.~(\ref{46}) we have assumed $ g_{\gamma,F\beta} =
g_{\gamma,B\beta} $ and $ g_{\gamma,\alpha F} = g_{\gamma,\alpha
B} $. The functions $ {\cal F}^{m,\rm surf} $ describing surface
contributions disappear in the limit $ L \rightarrow 0 $, i.e. the
surface contribution from the output boundary completely
compensates that from the input boundary. Comparison of the
expressions in Eqs.~(\ref{45}) and (\ref{46}) reveals a simple
relation between the volume and surface contributions:
\begin{eqnarray}       
 \frac{{\cal F}^{m,\rm surf}_{\gamma,\alpha\beta}(L,\omega_s,\omega_i) }{
  {\cal F}^{\rm vol}_{\gamma,\alpha\beta}(L,\omega_s,\omega_i) } &=& \frac{\Delta
  k_{\gamma,\alpha\beta}(\omega_s,\omega_i) }{ k_m(\omega_m) } \nonumber \\
 & \equiv & {\cal V}_{\gamma,\alpha\beta}^m(\omega_s,\omega_i) , \nonumber \\
 & & \hspace{-1cm} m=s,i; \hspace{5mm} \alpha,\beta,\gamma = F,B.
\label{47}
\end{eqnarray}
The structure of surface contribution is formed by mutual
interference of fields generated at the input and output
boundaries. At a boundary, only the energy conservation restricts
properties of an emitted photon pair. Such photon pair thus has a
rich internal spectral structure because phase-matching conditions
do not apply. It is then the mutual interference of fields coming
from the input and output boundaries that gives conditions similar
to those of phase matching naturally found in the volume
interaction.

In standard bulk sources of photon pairs that are typically
several mm long, the interaction among forward-propagating pump,
signal, and idler fields is important. In this case, the
commonly-used formalism for the description of SPDC [using a
two-photon spectral amplitude $ \Phi(\omega_s,\omega_i)] $ can be
applied (see, e.g., \cite{Keller1997,Grice1998,PerinaJr1999}). The
surface contributions can be involved in this formalism if the
following formal substitution is done:
\begin{eqnarray}   
 \Phi(\omega_s,\omega_i) & \longleftarrow & \sqrt{ 1+{\cal V}^{s}_{F,FF}
  (\omega_s,\omega_i) } \nonumber \\
 & &  \hspace{-5mm} \times \sqrt{ 1+{\cal V}^{i}_{F,FF}(\omega_s,\omega_i)
 } \, \Phi^{\rm vol}(\omega_s,\omega_i) ,
\label{48}
\end{eqnarray}
where the two-photon amplitude $ \Phi^{\rm vol}(\omega_s,\omega_i)
$ characterizes the usual volume contribution. Under the usual
condition of phase-matched nonlinear interaction [$ \Delta
k_{F,FF}(\omega_s^0,\omega_i^0) = 0 $] the surface contributions
occur only at spectral tails and are negligible.

\begin{figure}  
 \raisebox{4 cm}{a)} \hspace{0mm}
 \resizebox{0.85\hsize}{0.5\hsize}{\includegraphics{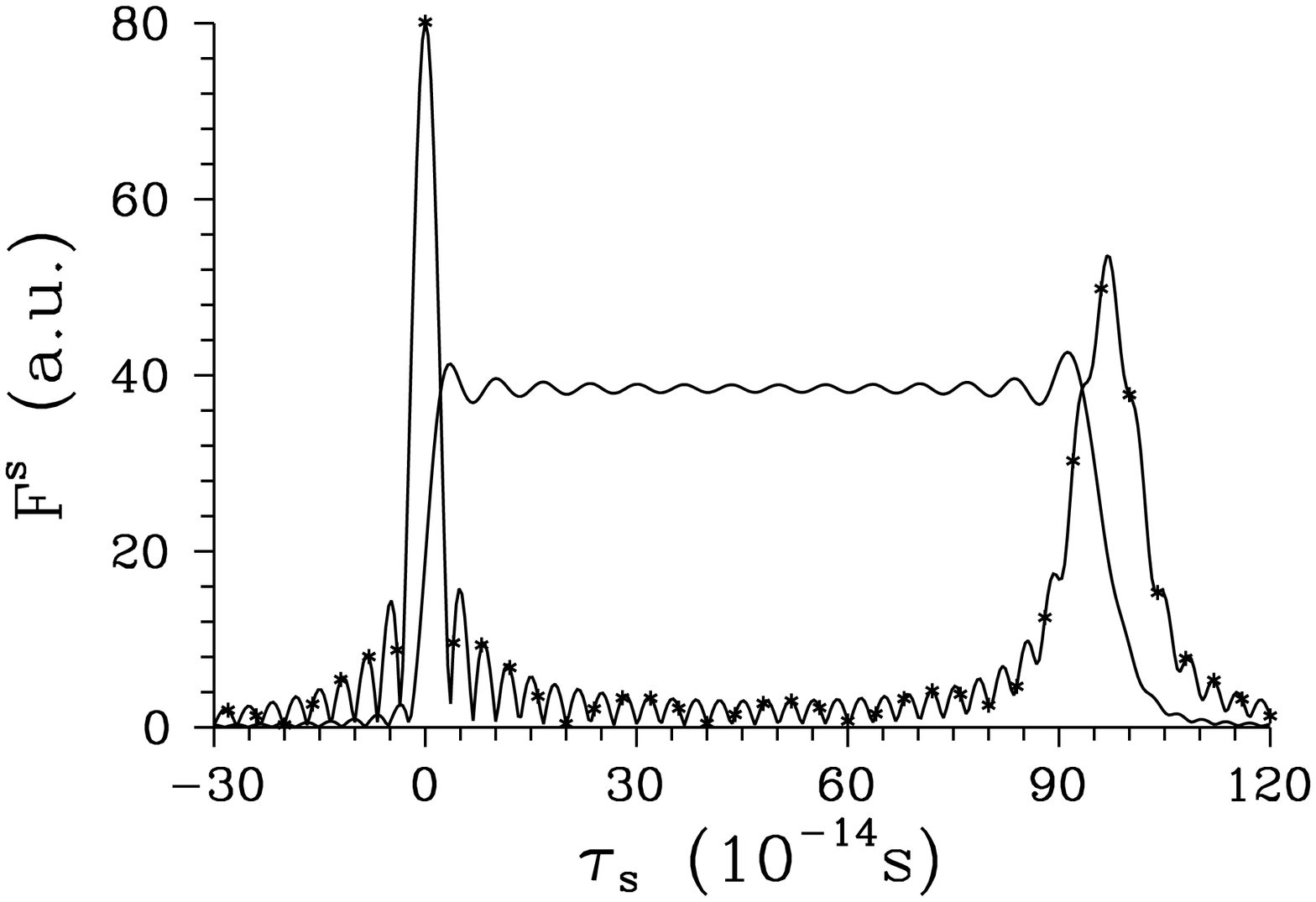}}

 \vspace{5mm}
 \raisebox{4 cm}{b)} \hspace{0mm}
 \resizebox{0.85\hsize}{0.5\hsize}{\includegraphics{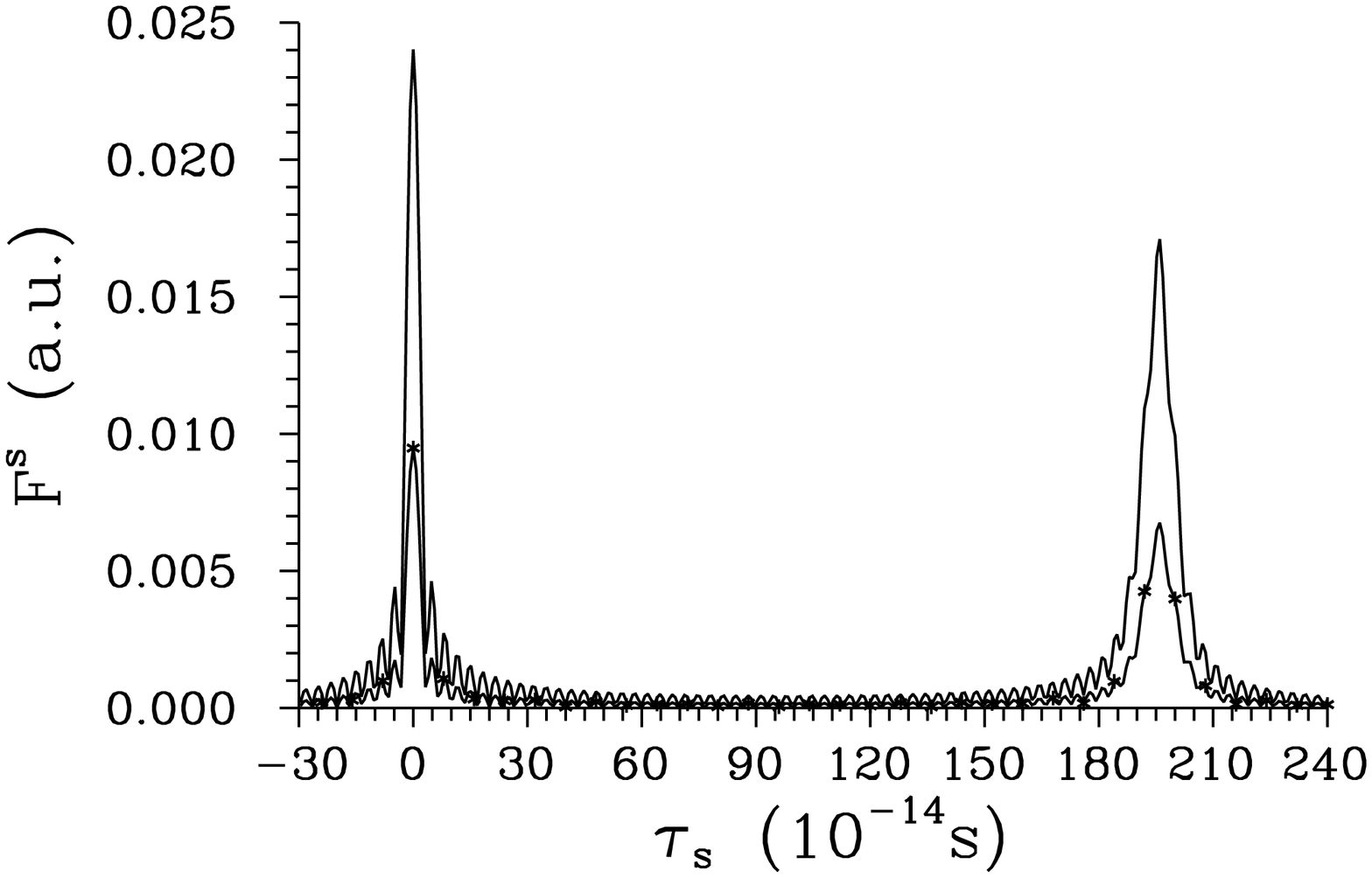}}
 \caption{Cross-section of the absolute value of two-photon temporal amplitude
  $ F^s(\tau_s) \equiv | \tilde{\cal F}^{b}_{F,FF}(\tau_s,0)| $ for the surface
  ($ b=\rm s,surf $, solid line with *) and volume ($ b=\rm vol $, solid line)
  contributions assuming cw pumping. The values of $ |\tilde{\cal F}^{s,\rm surf}_{F,FF}| $ are 5000
  (10) times magnified with respect to these of $ |\tilde{\cal F}^{\rm vol}_{F,FF}|
  $ in case a (b). The curves are appropriate for a BBO crystal 5~mm long and collinear type-II
  interaction at the pump wavelength of 400 nm and signal and idler wavelengths
  800~nm (frequency filters 30~nm wide (FWHM) are used). The crystal optical axis declines
  by 42.35~deg (perfect phase matching) (a) and 80~deg (b) with respect to the axis of fields'
  propagation.}
\label{fig2}
\end{figure}

Surface contributions have typically broader spectra compared to
the volume contributions. If these spectra are not filtered in an
experimental setup they lead to sharper features of two-photon
temporal amplitudes $ \tilde{\cal F}^{s,\rm surf}(\tau_s,\tau_i) $
and $ \tilde{\cal F}^{i,\rm surf}(\tau_s,\tau_i) $ [for their
definition, see Eq.~(\ref{56}) below]. As documented in Fig.
\ref{fig2}a for a BBO crystal, the surface two-photon temporal
amplitudes $ \tilde{\cal F}^{m,\rm surf} $ ($ m=s,i $) attain
large values in the vicinity of the input and output boundaries.
On the other hand the two-photon temporal amplitude $ \tilde{\cal
F}^{\rm vol} $ of the volume contribution has roughly the same
values along the whole nonlinear crystal. If the volume SPDC is
strongly phase mismatched its two-photon temporal amplitude $
\tilde{\cal F}^{\rm vol} $ resembles that of the surface SPDC (see
Fig.~\ref{fig2}b). This means that only values of the two-photon
temporal amplitude $ \tilde{\cal F}^{\rm vol} $ characterizing
photon pairs born in the vicinity of crystal edges are higher.
Destructive interference inside the nonlinear crystal prevails in
this case and suppresses photon-pair emission. Because the signal-
and idler-field spectra are very broad for phase-mismatched
interaction the shapes of two-photon temporal amplitudes $
\tilde{\cal F} $ for the volume and surface interactions are
similar. We also note that the temporal widths of peaks of
amplitudes $ F^s(\tau_s) $ in Fig.~\ref{fig2} are broader in the
area that corresponds to the beginning of the crystal compared to
those coming from the crystal end (and occurring around $ \tau_s
\approx 0 $~s) because of intermodal dispersion faced by a
photon-pair as it propagates through the crystal.

The fact that only photon pairs around the boundaries are
generated in the strongly phase-mismatched interaction resembles
the behavior of the second-harmonic field in the process of
strongly phase-mismatched second-harmonic generation
\cite{Centini2008}. Here, the pulsed second-harmonic field
propagates below the fundamental pulsed field (they have the same
group velocities) along the crystal and does not feel any
absorption \cite{Centini2008}. This can be interpreted so that the
second-harmonic field at the crystal output is generated only in
the vicinity of the output boundary.

\begin{figure}  
 \resizebox{0.9\hsize}{0.5\hsize}{\includegraphics{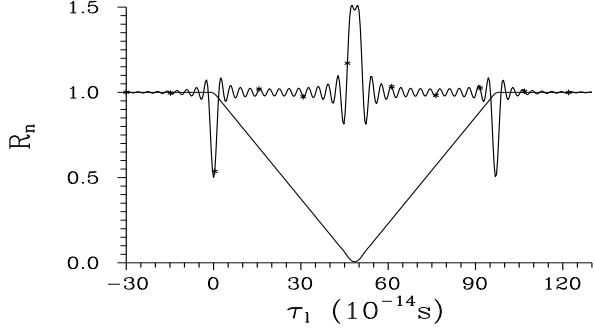}}
  \caption{Normalized coincidence-count rate $ R_n $ in Hong-Ou-Mandel
  interferometer as a function of
  relative time delay $ \tau_l $ is shown assuming volume
  (solid line) and surface (solid line with *) SPDC.
  Values of parameters appropriate for Fig.~2a are used.}
\label{fig3}
\end{figure}

Two-photon temporal amplitudes are not experimentally accessible
but certain information about their shape can be reached
\cite{PerinaJr1999} when measuring coincidence-count interference
rates $ R_n $ [defined in Eq.~(\ref{58}) below] in a
Hong-Ou-Mandel interferometer. Whereas a triangular dip is typical
for phase-matched volume SPDC, two side dips with reduced
visibility (around 0.5) and one central peak occur in the
coincidence-count rate $ R_n $ for surface SPDC as a consequence
of the shape of two-photon temporal amplitude with two peaks (see
Fig.~\ref{fig3}). We note that the profile of coincidence-count
rate $ R_n $ of strongly phase-mismatched volume SPDC is similar
to that appropriate for surface SPDC.
Alternatively, sum-frequency generation of the signal and idler
fields can be used to experimentally `scan' the shape of
two-photon temporal amplitude \cite{Harris2007,Nasr2008}. The
intensity field arising from sum-frequency generation should have
two peaks depending on the mutual temporal delay of the signal and
idler photons in the studied phase-mismatched case. These peaks
correspond to two boundaries.

Contrary to the case of usual nonlinear crystals, surface SPDC
cannot be neglected in nonlinear layered media (see Sec.~IV
later).

\section{Quantities characterizing the emitted photon pairs}

The inclusion of surface contributions to the expressions for
operators $ \hat{a}_{s_F} $, $ \hat{a}_{s_B} $, $ \hat{a}_{i_F} $,
and $ \hat{a}_{i_B} $ requires a generalization of the usually
used formulas for the determination of physical quantities
characterizing photon pairs. For simplicity we pay attention to
photon pairs with both photons propagating forward and define
operators $ \hat{a}_{m}(\omega_m) \equiv t_m(\omega_m)
\hat{a}_{m_F}(L,\omega_m) $ and $ \hat{E}^{(+)}_m(\tau_m) \equiv
t_m(\omega_m^0) \hat{E}^{(+)}_{m_F}(L,\tau_m) $, where the
coefficients $ t_m $ describe the amplitude transmissivities at
the output boundary ($ m=s,i $). We have $ t_m =
2n_m/(n_m+n_m^{(1)}) $ according to Fresnel's formulas, $ n_m $ ($
n_m^{(1)} $) means an index of refraction of (beyond) the
nonlinear medium ($ m=s,i $).

The joint signal-idler photon-number density $
n(\omega_s,\omega_i) $ giving the number of emitted photon pairs
with a signal photon in the unit interval around frequency $
\omega_s $ and its idler twin in the unit interval around
frequency $ \omega_i $ at the medium output plane is defined as
follows:
\begin{eqnarray}   
 n(\omega_s,\omega_i) &=& \left\langle
  \left[  \hat{a}_s^\dagger(\omega_s)
  \hat{a}_s(\omega_s) \hat{a}_i^\dagger (\omega_i)
  \hat{a}_i(\omega_i)  + {\rm h.c.}  \right]
  \right\rangle /2, \nonumber \\
 & &
\label{49}
\end{eqnarray}
Symbol $ \langle \, \rangle $ means averaging over the initial
signal- and idler-field vacuum states. Substitution of
Eqs.~(\ref{42}) and (\ref{43}) into the definition of the joint
signal-idler photon-number density $ n $ in Eq.~(\ref{49}) results
in the formula:
\begin{equation}  
 n(\omega_s,\omega_i) = {\rm Re} \{
  \tilde{\cal F}^{s*}(\omega_s,\omega_i)
  \tilde{\cal F}^{i}(\omega_s,\omega_i) \} .
\label{50}
\end{equation}
The functions $ \tilde{\cal F}^{s} $ and $ \tilde{\cal F}^{i} $ in
Eq.~(\ref{50}) include transmission through the output boundary:
\begin{equation} 
 \tilde{\cal F}^{m}(\omega_s,\omega_i) = t_s(\omega_s) t_i(\omega_i)
  {\cal F}^{m}_{F,FF}(\omega_s,\omega_i), \,\, m=s,i .
\label{51}
\end{equation}
In Eq.~(\ref{50}), symbol $ {\rm Re} $ denotes the real part of an
argument. We note that expressions containing factor $
\delta^2(\omega_p) $ occur in formulas like that written in
Eq.~(\ref{50}) for cw pumping. This factor has to be replaced
using the formula $ \delta^2(\omega_p) = 2T/(2\pi)
\delta(\omega_p) $ in this case where $ 2T $ means the
detection-interval length \cite{PerinaJr1999}.

Intensity spectrum $ S_s(\omega_s) $ of, e.g., the signal field
can be easily derived from the joint photon-number density $ n $;
\begin{eqnarray}   
 S_s(\omega_s) &=& \hbar\omega_s \int_{0}^{\infty} d\omega_i
  \, n(\omega_s,\omega_i) \nonumber \\
 &=& \hbar\omega_s \int_{0}^{\infty} d\omega_i
  \, {\rm Re} \{
  \tilde{\cal F}^{s*}(\omega_s,\omega_i)
  \tilde{\cal F}^{i}(\omega_s,\omega_i) \} . \nonumber \\
 & &
\label{52}
\end{eqnarray}

Photon flux of, e.g., the signal photons $ {\cal N}_s(\tau_s) $
beyond the nonlinear medium can be derived along the following
formula considering only photons emitted in photon pairs:
\begin{eqnarray}   
 {\cal N}_s(\tau_s) &=& \epsilon_0 c n_s^{(1)}(\omega_s^0) {\cal A} \int d\omega_i
   \nonumber \\
 & & \hspace{-2cm} \left\langle  \left[ \hat{E}_s^{(-)}(\tau_s) \hat{E}_s^{(+)}(\tau_s)
  \hat{a}_i^\dagger(\omega_i)\hat{a}_i(\omega_i)
  + {\rm h.c.} \right]  \right\rangle .
\label{53}
\end{eqnarray}
Using the solutions in Eqs.~(\ref{42}) and (\ref{43}) we arrive at
the expression:
\begin{eqnarray}   
 {\cal N}_s(\tau_s) &=& \frac{\hbar}{2\pi} \int d\omega_{s}
  \sqrt{\omega_{s}} \int d\omega'_{s} \sqrt{\omega'_{s}} \int d\omega_i
  \nonumber \\
  & & \hspace{-2.5cm} {\rm Re} \{ \exp[i(\omega_{s}-\omega'_{s})\tau_s]
   \tilde{\cal F}^{i*}(\omega_{s},\omega_i)
   \tilde{\cal F}^{s}(\omega'_{s},\omega_i) \} .
\label{54}
\end{eqnarray}
Assuming a narrow idler-field spectrum we can rearrange the
formula in Eq.~(\ref{54}) as follows:
\begin{equation}  
 {\cal N}_s(\tau_s) = \hbar \omega^0_s \int d\tau_i \, {\rm Re}
  \{ \tilde{\cal F}^{i*}(\tau_s,\tau_i) \tilde{\cal F}^{s}(\tau_s,\tau_i) \}.
\label{55}
\end{equation}
The functions $ \tilde{\cal F}^s $ and $ \tilde{\cal F}^i $ in
time domain in Eq.~(\ref{55}) are derived from their spectral
counterparts given in Eqs.~(\ref{51}), (\ref{45}), and (\ref{46})
along the formula:
\begin{eqnarray}   
 \tilde{\cal F}^m(\tau_s,\tau_i) &=& \frac{1}{2\pi} \int d\omega_s \int
  d\omega_i \, \sqrt{ \frac{\omega_s \omega_i}{\omega_s^0
  \omega_i^0} } \tilde{\cal F}^m(\omega_s,\omega_i)  \nonumber \\
 & & \hspace{-1cm}  \times  \exp(-i\omega_s \tau_s)
  \exp(-i\omega_i \tau_i) , \hspace{5mm} m=s,i .
\label{56}
\end{eqnarray}

The number $ N $ of coincidence counts caused by a simultaneous
detection of both photons from one pair in a Hong-Ou-Mandel
interferometer is given as:
\begin{eqnarray}   
 N(\tau_l) &=& \frac{ \epsilon_0^2 c^2
  n_s^{(1)}(\omega_s^0) n_i^{(1)}(\omega_i^0){\cal A}^2
  }{ 2 \hbar^2 \omega_s^0\omega_i^0 }
  \nonumber \\
 & & \hspace{-1cm} \times
  \int dt_1 \int dt_2 \left[ (|r|^4+|t|^4)
  \left\{ \langle \hat{E}_s^{(-)}(t_1)  \right. \right.
  \nonumber \\
 & & \hspace{-1cm} \left. \times \hat{E}_s^{(+)}(t_1)
  \hat{E}_i^{(-)}(t_2) \hat{E}_i^{(+)}(t_2) \rangle + {\rm c.c.}
  \right\} \nonumber \\
 & & \hspace{-1cm} + \left\{ (r^*t)^2 \langle \hat{E}_s^{(-)}(t_1) \hat{E}_s^{(+)}(t_2)
  \hat{E}_i^{(-)}(t_2-\tau_l) \right. \nonumber \\
 & & \hspace{-1cm} \left. \left. \times \hat{E}_i^{(+)}(t_1-\tau_l)
  \rangle + {\rm c.c.} \right\} \right] ;
\label{57}
\end{eqnarray}
$ r $ ($ t $) stands for amplitude reflectivity (transmissivity)
of a beam-splitter in the interferometer and $ \tau_l $ denotes a
relative time delay introduced between the signal and idler
photons. Using the solutions in Eqs. (\ref{42}) and (\ref{43}) the
normalized coincidence-count number $ R_n $ can be derived in the
form:
\begin{equation}   
 R_n(\tau_l) = 1- \rho(\tau_l) ,
\label{58}
\end{equation}
where
\begin{eqnarray}  
 \rho(\tau_l) &=&
  \frac{1}{4 R_0} \int d\omega_s \int d\omega_i
  \frac{\omega_s\omega_i}{\omega_s^0\omega_i^0}
  {\rm Re} \left\{ (r^*t)^2 \right. \nonumber \\
 & & \hspace{-1cm} \times \left.
  \tilde{\cal F}^{s*}(\omega_s,\omega_i) \tilde{\cal F}^{i}(\omega_i,\omega_s)
  \exp[i(\omega_s-\omega_i)\tau_l] \right\}
\label{59}
\end{eqnarray}
and
\begin{eqnarray}  
 R_0 &=& \frac{1}{8}
  ( |r|^4 + |t|^4 ) \int d\omega_s \int d\omega_i
  \frac{\omega_s\omega_i}{\omega_s^0\omega_i^0}
  \nonumber \\
 & & \hspace{0cm} \times {\rm Re} \left\{
  \tilde{\cal F}^{s*}(\omega_s,\omega_i) \tilde{\cal F}^{i}(\omega_s,\omega_i)
  \right\} .
\label{60}
\end{eqnarray}

\section{Nonlinear layered structures}

We now consider structures composed of both linear and nonlinear
layers with thicknesses typically in hundreds of nm. Fulfilment of
phase-matching conditions is not important in these short layers,
because $ \Delta k L \ll \pi $ ($ L $ means a typical layer
length). Surface SPDC becomes important in these structures
because volume contributions to SPDC from individual layers are
weak. A generalization of the developed theory to layered
structures is straightforward and is based on the fact that we
detect only one photon pair. This pair is generated in one of the
nonlinear layers and propagates to the output of the structure.
The theory giving volume contributions to SPDC has been developed
in \cite{PerinaJr2006,Centini2005,PerinaJr2007} using a
perturbation solution to Schr\"{o}dinger equation. Here, we
restrict ourselves to a scalar model and collinear interaction to
emphasize the important steps in the derivation. A generalization
including general directions of fields' propagation and their
polarization properties is straightforward following the way
presented in \cite{PerinaJr2006}.

First we generalize the momentum operator $ \hat{G}_{\rm int} $
written in Eq.~(\ref{3}) to include the nonlinear interaction in $
N $ layers:
\begin{eqnarray}   
 \hat{G}_{\rm int}(z) &=& \frac{4 \epsilon_0 {\cal A}}{ \sqrt{2\pi}}
  \int d\omega_p \int d\omega_s \int d\omega_i \nonumber \\
 & &
  \hspace{-15mm} \delta(\omega_p-\omega_s-\omega_i)
  \sum_{l=1}^{N} \sum_{\alpha,\beta,\gamma=F,B} d_{\gamma,\alpha\beta}^{(l)}
  \nonumber \\
 & & \hspace{-15mm} \times
  \left[ E^{(+,l)}_{p_\gamma}(z,\omega_p) \hat{E}^{(-,l)}_{s_\alpha}(z,\omega_s)
  \hat{E}^{(-,l)}_{i_\beta}(z,\omega_i)
  + {\rm h.c.} \right] , \nonumber \\
 & &
\label{61}
\end{eqnarray}
where the amplitude operators $ \hat{E}^{(-,l)}_{m_\alpha} $ are
defined in $ l $th layer:
\begin{eqnarray} 
 \hat{E}_{m_\alpha}^{(-,l)}(z,\omega_m) &=& - i \sqrt{ \frac{
  \hbar\omega_m }{ 2\epsilon_0 c{\cal A} n_m^{(l)}(\omega_m) } }
  {\rm rect}_{z_{l-1},z_l}(z) \nonumber \\
  & & \hspace{-15mm} \times \hat{a}^{(l)\dagger}_{m_\alpha}(z,\omega_m),
   \hspace{5mm} m = s,i, \hspace{5mm} \alpha=F,B.
\label{62}
\end{eqnarray}
The pump-field spectral amplitude $ E^{(+,l)}_{p_\gamma} $ ($
\gamma=F,B $) is defined inside the $ l $th layer similarly as the
amplitudes in Eq.~(\ref{62}). The nonlinear coefficients $
d_{\gamma,\alpha\beta}^{(l)} $ as well as indices of refraction $
n_m^{(l)}(\omega_m) $ characterize the $ l $th layer that extends
from $ z_{l-1} $ to $ z_l $. Function $ {\rm rect}_{a,b}(z) $
equals 1 for $ a<z<b $ and is zero otherwise.

Photon-pair generation in the $ l $th layer can be studied using
the approach and results presented in Sec.~II. These formulas give
us appropriate operators $ \hat{a}_{m_\alpha}(z_l,\omega_m) $ ($
m=s,i $; $ \alpha=F,B $) at the end of the $ l $th layer.
Fresnel's relations at the boundaries have to be used to
'transfer' a generated photon pair outside the boundaries of the
layered structure. A photon pair can be emitted in any of $ N $
layers and the corresponding quantum trajectories have to be
superposed. So we can write:
\begin{eqnarray}   
 \hat{a}_{m_\alpha}^{\rm out}(\omega_m) &=& \sum_{l=1}^{N} \sum_{\alpha'=F,B} {\cal
  T}^{m,(l)}_{\alpha\alpha'}(\omega_m) \hat{a}_{m_{\alpha'}}^{(l)}(z_l,\omega_m) ,
  \nonumber \\
 & & \hspace{10mm}  m=s,i, \;\; \alpha=F,B .
\label{63}
\end{eqnarray}
The coefficients $ T^{m,(l)}_{\alpha\alpha'} $ in Eq.~(\ref{63})
can be derived using the propagation matrix method and Fresnel's
relations at boundaries \cite{PerinaJr2006,Yeh1988}.

Properties of photon pairs as described by quantities introduced
in Sec.~III and measured by a simultaneous detection of both
photons comprising a photon pair can be derived from the
fourth-order correlation function $ \langle
\hat{a}_{s_{\alpha'}}^{{\rm out}\dagger}(\omega'_s)
\hat{a}_{s_{\alpha}}^{{\rm out}}(\omega_s)
\hat{a}_{i_{\beta'}}^{{\rm out}\dagger}(\omega'_i)
\hat{a}_{i_{\beta}}^{{\rm out}}(\omega_i) \rangle $. Substituting
Eq.~(\ref{63}) into the definition of the correlation function and
using the solution for one nonlinear layer as presented in
Eqs.~(\ref{42}) and (\ref{43}) we arrive at:
\begin{eqnarray}   
 \langle \hat{a}_{s_{m\alpha'}}^{{\rm out}\dagger}(\omega'_s)
  \hat{a}_{s_{\alpha}}^{{\rm out}}(\omega_s) \hat{a}_{i_{\beta'}}^{{\rm
  out}\dagger}(\omega'_i) \hat{a}_{i_{\beta}}^{{\rm out}}(\omega_i)
  \rangle &=& \nonumber \\
 & & \hspace{-3cm}  {\cal F}^{s,{\rm out} *}_{\alpha'\beta'}(\omega'_s,\omega'_i)
  {\cal F}^{i,{\rm out}}_{\alpha\beta}(\omega_s,\omega_i) , \nonumber \\
 & & \hspace{-2cm}
  \alpha,\alpha',\beta,\beta'=F,B .
\label{64}
\end{eqnarray}
The two-photon amplitudes $ {\cal F}^{o,{\rm out}}_{\alpha\beta} $
($ o=s,i $) describing a photon pair at the output boundaries with
the photons propagating in directions indicated by lower indices $
\alpha $ and $ \beta $ can be expressed in terms of two-photon
amplitudes $ {\cal F}^{o,(l)}_{\alpha'\beta'} $ characterizing
individual layers:
\begin{eqnarray}   
 {\cal F}^{o,{\rm out}}_{\alpha\beta}(\omega_s,\omega_i) &=& \sum_{l=1}^{N}
  \sum_{\alpha',\beta'=F,B} {\cal T}^{s,(l)}_{\alpha\alpha'}(\omega_s)
  {\cal T}^{i,(l)}_{\beta\beta'}(\omega_i)  \nonumber \\
 & & \hspace{-2cm} \times {\cal
  F}^{o,(l)}_{\alpha'\beta'}(\omega_s,\omega_i), \hspace{5mm} o=s,i;
  \hspace{5mm} \alpha,\beta=F,B .
\label{65}
\end{eqnarray}
Expressions in Eqs.~(\ref{45}) and (\ref{47}) can be rearranged
into the form:
\begin{eqnarray}   
 {\cal F}^{o,(l)}_{\alpha\beta}(\omega_s,\omega_i) &=&
   \sum_{\gamma=F,B} g^{(l)}_{\gamma,\alpha\beta}(\omega_s,\omega_i) \left[ 1+ {\cal
  V}^{o,(l)}_{\gamma,\alpha\beta}(\omega_s,\omega_i) \right]
   \nonumber \\
 & & \hspace{-1.5cm} \times  E^{(+,l)}_{p_\gamma}(\omega_s+\omega_i)
   \exp[ik^{(l)}_{p_\gamma}(\omega_s+\omega_i)L_l] \nonumber \\
 & & \hspace{-1.5cm} \times
   \exp[-i\Delta k^{(l)}_{\gamma,\alpha\beta}(\omega_s,\omega_i)
   L_l/2] \nonumber \\
 & & \hspace{-1.5cm} \times  L_l \,
   {\rm sinc} [\Delta k^{(l)}_{\gamma,\alpha\beta}(\omega_s,\omega_i) L_l/2] ,
   \nonumber \\
 & & \hspace{1cm}  o=s,i; \hspace{5mm} \alpha,\beta=F,B,
\label{66}
\end{eqnarray}
where
\begin{eqnarray}   
 {\cal V}_{\gamma,\alpha\beta}^{o,(l)} (\omega_s,\omega_i) &=& \frac{\Delta
  k^{(l)}_{\gamma,\alpha\beta}(\omega_s,\omega_i) }{ k^{(l)}_o(\omega_o) } ,
  \nonumber \\
  & & o=s,i; \hspace{5mm} \alpha,\beta,\gamma= F,B
\label{67}
\end{eqnarray}
and $ g_{\gamma,\alpha\beta}^{(l)}(\omega_s,\omega_i) = 2i
d_{\gamma,\alpha\beta}^{(l)} \sqrt{ \omega_s \omega_i} / [2\pi c
\sqrt{ n_s^{(l)}(\omega_s) n_i^{(l)}(\omega_i)} ] $. Superscript $
(l) $ denotes quantities appropriate for the $ l $th layer and $
L_l $ means the length of this layer ($ L_l=z_l-z_{l-1} $). Symbol
$ E^{(+,l)}_{p_\gamma} $ occurring in Eq.~(\ref{66}) refers to the
spectral amplitude of field $ p_\gamma $ ($ \gamma=F,B $) at the
end of the $ l $th layer (at $ z = z_l $) and can be determined
using the propagation matrix formalism. In deriving Eq.~(\ref{66})
we have assumed that $ g^{(l)}_{\gamma,F\beta} =
g^{(l)}_{\gamma,B\beta} $ and $ g^{(l)}_{\gamma,\alpha F} =
g^{(l)}_{\gamma,\alpha B} $.

The signal-field intensity spectrum $ {\cal S}_s $, its photon
flux $ {\cal N}_s $, and coincidence-count interference rate $ R_n
$ in the Hong-Ou-Mandel interferometer can then be determined
using the formulas in Eqs.~(\ref{52}), (\ref{54}), and
(\ref{58}---\ref{60}) using the spectral two-photon amplitudes $
{\cal F}^{s,{\rm out}} $ and $ {\cal F}^{i,{\rm out}} $ defined in
Eq.~(\ref{65}).

The generalization including fields' propagation under nonzero
angles of incidence is straightforward following the procedure
presented above in Sec.~II. The key point here is that the
necessary amplitude corrections assuring the fulfilment of
continuity requirements for the electric and magnetic fields at
the boundaries are defined only at one side of the boundaries. To
be more specific, we use nonzero amplitude-operator corrections $
\delta\hat{a}_{s_F}(0,\omega_s) $ and $
\delta\hat{a}_{s_B}(0,\omega_s) $ [$
\delta\hat{a}_{s_F}(L,\omega_s) $ and $
\delta\hat{a}_{s_B}(L,\omega_s) $] for the left-hand [right-hand]
side of the nonlinear medium (layer) to describe surface SPDC.
When nonzero angles of incidence are considered, we can argue as
follows. The angle of incidence of a given field equals (up to the
sign) the angle of reflection and so their cosines giving
multiplicative factors for projections of electric- and
magnetic-field amplitudes (lying in the plane of incidence) to the
plane of boundary are the same. This means that the equations
written in Eqs.~(\ref{28}), (\ref{29}), (\ref{38}), and (\ref{39})
remain valid for any angle of incidence. Equations~(\ref{66}) and
(\ref{67}) can then be used in this case provided that we use the
$ z $ components of wave vectors $ k_p $, $ k_s $, and $ k_i $
instead of their full lengths.

\begin{figure}  
 \resizebox{1.\hsize}{!}{\includegraphics{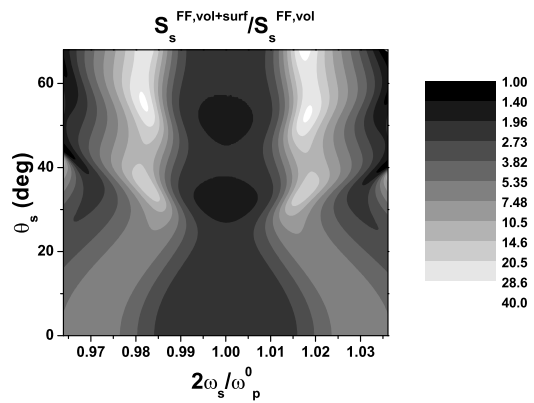}}
  \caption{Contour plot of the ratio $ S_s^{FF,\rm vol+surf}/S_s^{FF,\rm vol} $ of
  the signal-field spectra with ($ S_s^{FF,\rm vol+surf} $) and
  without ($ S_s^{FF,\rm vol} $) surface
  contributions to SPDC as it
  depends on signal-field emission angle $ \theta_s $. Both
  photons propagate forward. Logarithmic
  scale is used on the $ z $ axis.}
\label{fig4}
\end{figure}

Surface SPDC can significantly contribute to the number of the
generated photon pairs in layered structures. As an example, we
consider a structure composed of 25 layers of nonlinear GaN of the
thickness of 117~nm that sandwich 24 linear layers of AlN of the
thickness of 180~nm. This structure as a source of photon pairs
has been studied in detail in \cite{Centini2005,PerinaJr2006} for
s-polarized pump (normally-incident), signal, and idler beams and
the pump wavelength of 664.5~nm. Considering only volume
contributions efficient photon-pair generation occurs at
degenerate frequencies of the signal and idler fields for the
signal-field emission angle of 14~deg (see Fig.~5 in
\cite{PerinaJr2006}). The surface contributions lead to
additional non-negligible photon-pair generation, as documented in
Fig.~\ref{fig4} where the ratio $ S_s^{FF,\rm
vol+surf}/S_s^{FF,\rm vol} $ of signal-field spectral intensities
with ($ S_s^{FF,\rm vol+surf} $) and without ($ S_s^{FF,\rm vol}
$) surface contributions is plotted as a function of signal-field
emission angle $ \theta_s $.
This ratio is minimum under the conditions giving the best
constructive interference (i.e., when $ \omega_s = \omega_p^0/2 $
and $ \theta_s = 14~{\rm deg} $) and equals approximately 2. This
means that the surface contributions roughly double the number of
emitted photon pairs. It should be emphasized that constructive
interference between the volume and surface contributions plays
the key role here. Comparison of the graph in Fig.~\ref{fig4} with
that plotted in Fig.~5 in \cite{PerinaJr2006} indicates that the
worse the constructive interference inside the layered structure
the greater the relative contribution of surface terms. However,
the overall number of generated photon pairs is quite low in this
region.

\begin{figure}  
 \resizebox{0.9\hsize}{!}{\includegraphics{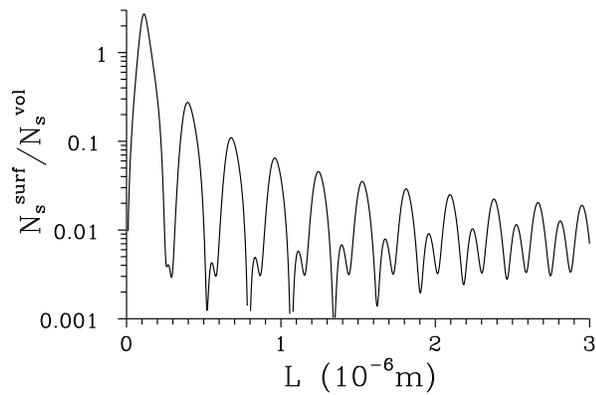}}
  \caption{Ratio $ N^{FF,\rm surf}/N^{FF,\rm vol} $ of
  the numbers of emitted photon pairs originating in surface
  ($ N^{FF,\rm surf} $) and volume ($ N^{FF,\rm vol} $) down-conversion from
  one layer of GaN of length $ L $ pumped at $ \lambda_p = 664.5 $~nm at normal
  incidence. Pairs with frequencies in the interval $ \Delta \Omega
  $ used in the graph in Fig.~\ref{fig4} and propagating along the $ +z $ axis are
  collected; $ N = \int_{\Delta\Omega} d\omega_s \int_{\Delta\Omega}
  d\omega_i \, n(\omega_s,\omega_i) $ and $ n $ is given
  in Eq.~(\ref{49}). Logarithmic scale on the $ y $ axis is used.}
\label{fig5}
\end{figure}

The increase of photon-pair generation rate is caused mainly by
processes that do not (even roughly) obey phase matching
conditions. Weights of the surface contributions for different
processes can be judged according to the value of parameter $
{\cal V}^{m,(l)}_{\gamma,\alpha\beta} $ ($ m=s,i $; $
\alpha,\beta,\gamma=F,B $) defined in Eq.~(\ref{67}). The
following values are met in GaN for the studied structure around
the point where the best constructive interference has been found:
$ {\cal V}^{m}_{F,FF} = - {\cal V}^{m}_{B,BB} \approx 0.05 $, $
{\cal V}^{m}_{F,FB} \approx {\cal V}^{m}_{F,BF} \approx - {\cal
V}^{m}_{B,FB} \approx {\cal V}^{m}_{B,BF} \approx 2 $, $ {\cal
V}^{m}_{F,BB} = - {\cal V}^{m}_{B,FF} \approx 4 $. Because the
dominant role in surface SPDC is played by the highly
phase-mismatched processes, lengths of nonlinear layers have to be
less or comparable to the coherence length of the nonlinear
interaction to observe these contributions. The ratio $ N^{FF,{\rm
surf}}/N^{FF,{\rm vol}} $ of the numbers of emitted photon pairs
in a certain spectral region coming from the surface ($ N^{FF,{\rm
surf}} $) and volume ($ N^{FF,{\rm vol}} $) contributions for one
layer of GaN as plotted in Fig.~\ref{fig5} indicates that the
inclusion of surface contributions is important for the lengths
below 1 $\mu $m.

\begin{figure}  
 \resizebox{0.9\hsize}{0.5\hsize}{\includegraphics{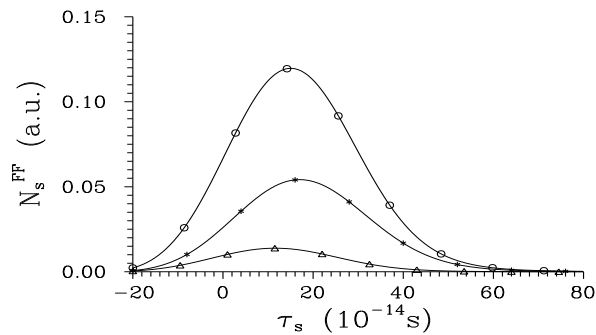}}
  \caption{Signal-field photon flux $ {\cal N}^{FF}_s $ including volume
  (solid line with *), surface (solid line with
  $ \triangle $), and volume + surface (solid line with $ \circ $)
  contributions. The structure is pumped by a Gaussian pulse with
  the duration of 250~fs (for details, see
  \cite{PerinaJr2006})}
\label{fig6}
\end{figure}

\begin{figure}  
 \resizebox{0.9\hsize}{0.5\hsize}{\includegraphics{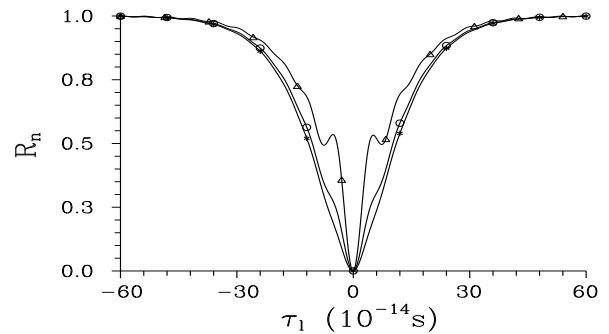}}
  \caption{Normalized coincidence-count rate $ R_n $ in Hong-Ou-Mandel
  interferometer as a function of
  relative time delay $ \tau_l $ is shown provided that volume
  (solid line with *), surface (solid line with
  $ \triangle $), and volume + surface (solid line with $ \circ $)
  contributions are included. The structure is pumped by
  a Gaussian pulse 250~fs long.}
\label{fig7}
\end{figure}

In general, the inclusion of surface contributions leads to the
broadening of spectral two-photon amplitudes $ {\cal F}^{s,{\rm
out}}(\omega_s,\omega_i) $ and $ {\cal F}^{i,{\rm
out}}(\omega_s,\omega_i) $. Thus widths of the signal- and
idler-field intensity spectra also increase. The Schmidt
decomposition of two-photon spectral amplitudes reveals that
entropy of its coefficients increases provided that the surface
terms are included. This means that entanglement between the
signal and idler fields increases due to surface SPDC. On the
other hand and considering pulsed pumping, photon fluxes $ {\cal
N} $ and coincidence-count interference rates $ R_n $ in a
Hong-Ou-Mandel interferometer as temporal characteristics become
narrower as documented in Figs. \ref{fig6} and \ref{fig7}. The
overall photon flux $ {\cal N} $ of, e.g., the pulsed signal field
occurs earlier at the output of the nonlinear structure compared
to the case of only the volume interaction because the structure
is 'less-resonant' ('less-transparent') for the surface
contributions than for the volume ones. Phase modulation of the
two-photon spectral amplitude characterizing the surface
contributions provides the coincidence-count pattern $ R_n(\tau_l)
$ in the form of a global dip with visibility equal to 1 and two
small side-dips (see Fig.~\ref{fig7}).

Surface effects in SPDC occur at discontinuities of $ \chi^{(2)} $
nonlinearity. Such discontinuities are found also in
periodically-poled nonlinear crystals like LiNbO$ {}_3 $. In these
crystals, an enhancement of photon-pair generation rates is
expected provided that the period of poling is sufficiently short
so that a sufficienly large number of discontinuities is present
inside the nonlinear crystal. The surface effects are also
expected in wave-guiding geometries where the conditions of total
reflection of fields at the boundaries are met. This might be
important mainly in photonic fibers. The studied surface effects
are by no means restricted to 1D geometries, even stronger surface
contributions might arise from 2D or 3D nonlinear structures. An
effective enhancement of the nonlinear interaction caused by the
surface effects should also be observed in stimulated $ \chi^{(2)}
$ processes like second-harmonic generation.

\section{Conclusions}

Surface spontaneous parametric down-conversion has been predicted
combining the solution of quantum Heisenberg equations and the
continuity requirements at boundaries. Formulas for the
determination of the number of generated photon pairs, spectra,
and photon-fluxes of the down-converted fields as well as
coincidence-count rates in a Hong-Ou-Mandel interferometer have
been derived using generalized signal- and idler-field two-photon
spectral amplitudes. It has been shown that surface contributions
from the input and output boundaries of a nonlinear crystal give
structures similar to those characterizing the volume
contributions. Nevertheless they are weak. The surface
contributions are important whenever strongly phase-mismatched
nonlinear interactions give considerable contributions, e.g., in
nonlinear layered structures. An example of a GaN/AlN structure
composed of several tens of layers has shown that the surface and
volume contributions can be comparable in their amplitudes. This
shows that the role of surface effects in other nonlinear
structures like periodically-poled materials, nonlinear
wave-guiding structures or structures with stimulated processes
should be addressed.

\acknowledgments{Support by projects IAA100100713 of GA AV \v{C}R,
MSM6198959213, COST 09026, 1M06002, and AVOZ 10100522 of the Czech
Ministry of Education is acknowledged. J.P. and O.H. thank M.
Scalora and E. Fazio for discussion about surface second-harmonic
generation.}


\begin{thebibliography}{12}

\bibitem{Bloembergen1962} N. Bloembergen and P.S. Pershan, Phys.
 Rev. {\bf 128}, 606 (1962).
\bibitem{Bloembergen1969} N. Bloembergen, H.J. Simon, and C.H. Lee, Phys.
 Rev. {\bf 181}, 1261 (1969).
\bibitem{Mlejnek1999} M. Mlejnek, E.M. Wright, J.V. Moloney, and
 N. Bloembergen, Phys. Rev. Lett. {\bf 83}, 2934 (1999).
\bibitem{Roppo2007} V. Roppo, M. Centini, C. Sibilia, M. Bertolotti, D. de Ceglia,
 M. Scalora, N. Akozbek, M.J. Bloemer, J.W. Haus, O.G. Kosareva, and V.P.
 Kandidov, Phys. Rev. A {\bf 76}, 033829 (2007).
\bibitem{Centini2008} M. Centini, V. Roppo, E. Fazio, F. Pettazzi, C. Sibilia, J.W. Haus, J.V.
 Foreman, N. Akozbek, M.J. Bloemer, and M. Scalora, Phys. Rev. Lett. {\bf 101}, 113905 (2008).
\bibitem{Mendoza1996} B.S. Mendoza and W.L. Moch\'{a}n, Phys. Rev.
 B {\bf 53}, 4999 (1996).
\bibitem{Mendoza1998} B.S. Mendoza, A. Gaggiotti, and R. Del Sole,
 Phys. Rev. Lett. {\bf 81}, 3781 (1998).
\bibitem{Altewischer2002} E. Altewischer, M.F. van Exter, and J.P.
 Woerdman, Nature {\bf 418}, 304 (2002).
\bibitem{Mandel1995} L. Mandel and E. Wolf,
 {\it Optical Coherence and Quantum Optics}
 (Cambridge University Press, Cambridge, 1995), chap. 22.4.
\bibitem{Boyd2003} R.W. Boyd, {\it Nonlinear Optics} (Academic
 Press, Amsterdam, 2003), 2nd edition.
\bibitem{Huttner1990} B. Huttner, S. Serulnik, Y. Ben-Aryeh, Phys.
 Rev. A {\bf 42}, 5594 (1990).
\bibitem{Luks2002} A. Luk\v{s} and V. Pe\v{r}inov\'{a}, {\it
 Progress in Optics} Vol. 43, Ed. E. Wolf (Elsevier, Amsterdam,
 2002), p. 295.
\bibitem{Vogel2001}  W. Vogel, D.G.Welsch, and S. Walentowicz, {\it
 Quantum Optics} (Wiley-VCH, Weinheim, 2001).
\bibitem{PerinaJr2009} J. Pe\v{r}ina Jr., A. Luk\v{s}, O. Haderka,
 and M. Scalora, Phys. Rev. Lett. {\bf 103}, 063902 (2009).
\bibitem{PerinaJr2000}  J. Pe\v{r}ina Jr. and J. Pe\v{r}ina, in {\it
 Progress in Optics} Vol. 41, Ed. E. Wolf (Elsevier, Amsterdam, 2000), p. 361.
\bibitem{Perina1991} J. Pe\v{r}ina, {\it Quantum Statistics of Linear and
 Nonlinear Optical Phenomena} (Kluwer, Dordrecht, 1991).
\bibitem{Keller1997} T. E. Keller and M. H. Rubin, Phys. Rev.
 A \textbf{56}, 1534 (1997).
\bibitem{Grice1998} W. P. Grice, R. Erdmann, I. A. Walmsley,
 D. Branning, Phys. Rev. A \textbf{57}, R2289 (1998).
\bibitem{PerinaJr1999} J. Pe\v{r}ina Jr., A. V. Sergienko, B. M. Jost,
 B. E. A. Saleh, M. C. Teich, Phys. Rev. A \textbf{59}, 2359
 (1999).
\bibitem{Harris2007} S. E. Harris, Phys. Rev. Lett. {\bf 98}, 063602 (2007).
\bibitem{Nasr2008} M. B. Nasr, S. Carrasco, B. E. A. Saleh, A. V.
 Sergienko, M. C. Teich, J. P. Torres, L. Torner, D. S. Hum, M. M.
 Fejer, Phys. Rev. Lett. {\bf 100}, 183601 (2008).
\bibitem{Wolf1980} M. Born and E. Wolf, {\it Principles of
 Optics} (Pergamon Press, Oxford, 1980), 6th edition.
\bibitem{PerinaJr2006} J. Pe\v{r}ina Jr., M. Centini, C. Sibilia, M. Bertolotti, and
 M. Scalora, Phys. Rev. A {\bf 73}, 033823 (2006).
\bibitem{Centini2005} M. Centini, J. Pe\v{r}ina Jr., L. Sciscione, C. Sibilia,
 M. Scalora, M.J. Bloemer, and M. Bertolotti, Phys. Rev. A {\bf
 72}, 033806 (2005).
\bibitem{PerinaJr2007} J. Pe\v{r}ina Jr., M. Centini, C. Sibilia, M. Bertolotti,
 and M. Scalora, Phys. Rev. A {\bf 75}, 013805 (2007).
\bibitem{Yeh1988} P. Yeh, {\it Optical Waves in Layered Media} (Wiley, New York, 1988).

\end{thebibliography}
\end{document}